\documentclass[letterpaper]{ieeeconf}

\usepackage{cite}
\usepackage{array}

\usepackage{amsthm}
\usepackage{amssymb}
\newcounter{thm}

\newtheorem{prob}[thm]{Problem}

\usepackage[pdftex]{graphicx}
\DeclareGraphicsExtensions{.pdf,.jpeg,.png}
\usepackage{url}
\usepackage{booktabs}
\usepackage{lettrine}

\usepackage{tikz,pgfplots}

\usetikzlibrary{shapes,arrows,positioning}
\usetikzlibrary{calc}
\usetikzlibrary{plotmarks}
\usepackage{fixmath}
\usepackage{amsmath}
\usepackage{amssymb}
\usepackage{mathrsfs}
\usepackage{gensymb}

\usepackage{algorithm}
\usepackage[noend]{algpseudocode}

\makeatletter
\def\BState{\State\hskip-\ALG@thistlm}
\makeatother

\floatname{algorithm}{Algorithm}

\usepackage{units}
\usepackage{mathtools}
\usepackage{dsfont}
\usepackage{booktabs}
\usepackage{bbm}
\usepackage{soul}
\IEEEoverridecommandlockouts

\newif\ifmargincomments 
\margincommentstrue

\ifmargincomments

\else

\fi

\newif\ifextendedversion 
\extendedversiontrue

\begin{document}
%
\title{\bf \LARGE Energy-optimal Design and Control of Electric Vehicles' Transmissions}

\author{Juriaan van den Hurk and Mauro Salazar
	\thanks{The authors are with the Control Systems Technology section, Eindhoven University of Technology (TU/e), Eindhoven, 5600 MB, The Netherlands, e-mail: {\tt\footnotesize j.l.j.v.d.hurk@student.tue.nl, m.r.u.salazar@tue.nl}}
}


%


\maketitle

\begin{abstract}
This paper presents models and optimization algorithms to jointly optimize the design and control of the transmission of electric vehicles equipped with one central electric motor (EM).
First,  considering the required traction power to be given, we identify a convex speed-dependent loss model for the EM.
Second, we leverage such a model to devise computationally-efficient algorithms to determine the energy-optimal design and control strategies for the transmission.
In particular, with the objective of minimizing the EM energy consumption on a given drive cycle, we analytically compute the optimal gear-ratio trajectory for a continuously variable transmission (CVT) and the optimal gear-ratio design for a fixed-gear transmission (FGT) in closed form, whilst efficiently solving the combinatorial joint gear-ratio design and control problem for a multiple-gear transmission (MGT), combining convex analysis and dynamic programming in an iterative fashion.
Third, we validate our models with nonlinear simulations, and benchmark the optimality of our methods with mixed-integer quadratic programming.
Finally, we showcase our framework in a case-study for a family vehicle, whereby we leverage the computational efficiency of our methods to jointly optimize the EM size via exhaustive search.
Our numerical results show that by using a 2-speed MGT, the energy consumption of an electric vehicle can be reduced by 3\% and 2.5\% compared to an FGT and a CVT, respectively, whilst further increasing the number of gears may even be detrimental due to the additional weight.
\end{abstract}


%
\IEEEpeerreviewmaketitle

\section{Introduction}
In the global shift away from fossil fuels, the transport sector is transitioning towards electrified powertrains, with electric vehicles (EVs) showing a great potential to replace conventional vehicles as their sales are substantially increasing every year~\cite{Un-NoorPadmanabanEtAl2017,IEA2020}.
However, a major barrier for the adoption of battery electric vehicles (BEVs) is the concern about their range and cost~\cite{FrankeGuentherEtAl2017,EgbueLong2012}. Thereby, increasing the energy efficiency of BEVs will not only improve their range, but also enable components' downsizing and cost reduction, therefore contributing to accelerating their adoption.

\begin{figure}
    \centering
    \includegraphics[width = 0.8\linewidth]{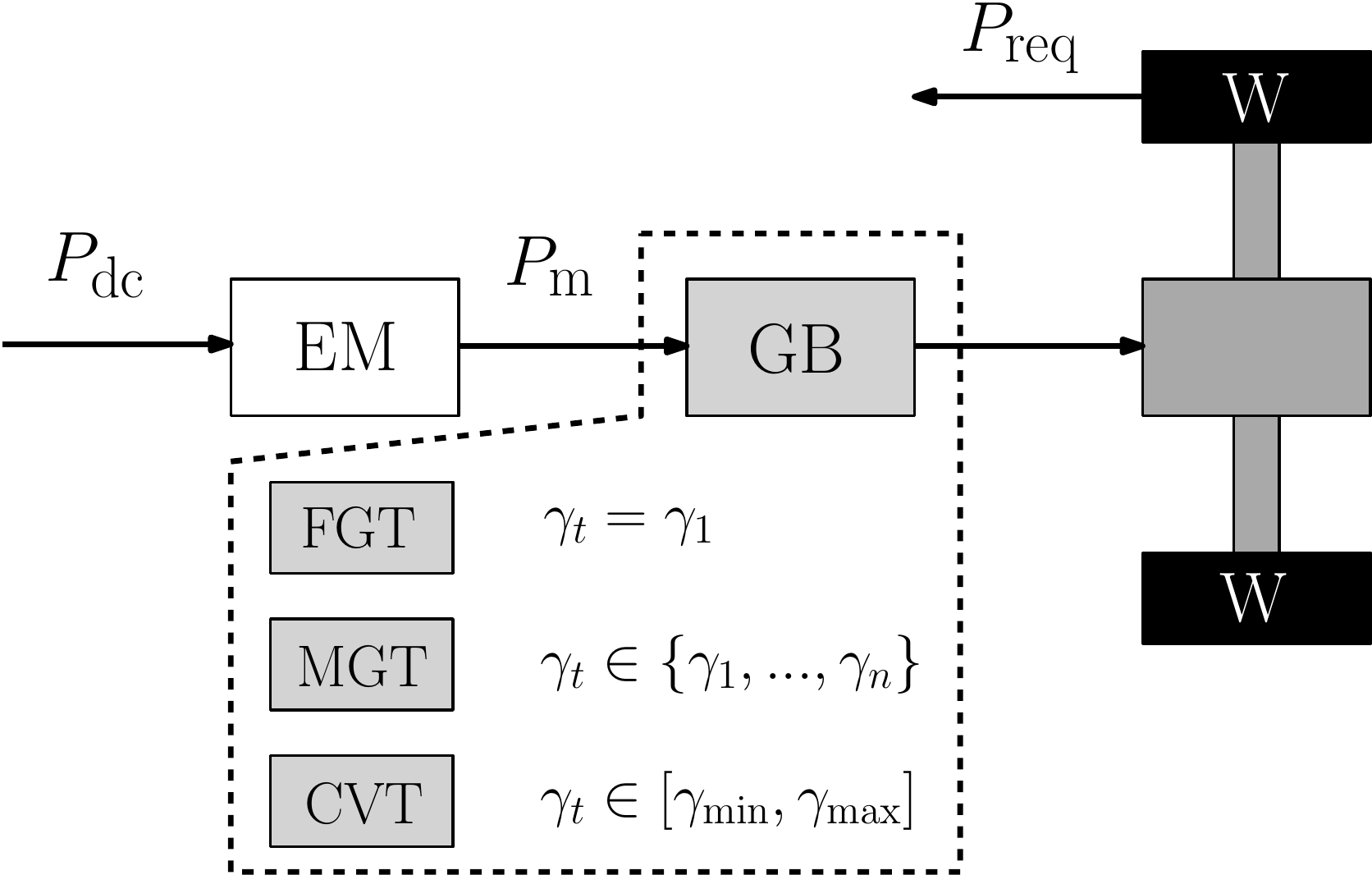}
    \caption{Schematic representation of the electric vehicle under consideration. It consists of an electric motor (EM), a transmission (GB) and the wheels (W). The arrows indicate the power flows between the components.}
    \label{fig:Veh_Model}
\end{figure}

Improving the energy consumption of EVs is a major technological challenge.
To decrease the mechanical energy demand of the vehicle, the shape of the body has been made more aerodynamic to reduce drag, whilst the mass of the components has been brought down.
To increase the energy efficiency of the powertrain itself, the design of its components should be optimized accounting for the control strategies used~\cite{Un-NoorPadmanabanEtAl2017}.
In this paper, we focus on the optimization of the design and control of the transmission of the EV shown in Fig.~\ref{fig:Veh_Model}, considering a fixed-gear transmission (FGT), a multiple-gear transmission (MGT) or a continuously variable transmission (CVT), whereby we capture the possibility of optimizing the electric motor (EM) size as well.

\subsubsection*{Related literature}
This work is related to two research streams.
In the first one, a number of strategies have been employed to optimize the gear-shift control of (hybrid) EVs. In~\cite{SainiSinghEtAl2016} a genetic algorithm is used to optimize the shifting patterns for an EV, whilst Pontryagin's principle has been used in~\cite{RitzmannChristonEtAl2019} to find the optimal gear selection and power distribution for a hybrid electric vehicle (HEV). The combination of dynamic programming (DP) and Pontryagin's principle has been successfully applied to find the optimal control of the gearshift command in HEVs~\cite{NgoHofmanEtAl2012}. Similarly, a combination of convex optimization and DP was employed to find the optimal gearshift command and power split for hybrid electric passenger~\cite{NueeschElbertEtAl2014,RobuschiSalazarEtAl2020} and race vehicles~\cite{DuhrChristodoulouEtAl2020}.
However, none of these techniques involve the optimization of the design of the transmission.

The second research line is concerned with the design of (hybrid) electric powertrains. The components' size and the energy management strategy of an HEV has been optimized jointly, leveraging convex optimization~\cite{MurgovskiJohannessonEtAl2012} and particle swarm optimization~\cite{EbbesenDoenitzEtAl2012}, but without accounting for the transmission.
Focusing on EVs, the transmission design and control problem has been studied analytically~\cite{HofmanSalazar2020}, and solved simultaneously using derivative-free methods~\cite{HofmanJanssen2017} and convex optimization~\cite{VerbruggenSalazarEtAl2019}, which was also applied to e-racing in~\cite{BorsboomFahdzyanaEtAl2021,LocatelloKondaEtAl2020}. However, all these approaches only consider FGTs or CVTs.
One of the challenges of finding the optimal gear-ratios and gear-shifting strategy for an MGT is the combinatorial nature of the gear-shifting problem, and the fact that the problems of finding the optimal shifting patterns and the optimal ratios are intimately coupled.
In~\cite{LeiseSimonEtAl2019}, mixed-integer nonlinear programming was used to find the optimal gear-ratios for an MGT, jointly computing the optimal gear-shifting control strategy.
The methods proposed in~\cite{SorniottiSubramanyanEtAl2011} compute the optimal gear-ratios for a given gear-shift map, whilst the authors of~\cite{GaoLiangEtAl2015} leverage DP to find the optimal gear-shift trajectory for a set of possible gear-ratios. In both contributions, the missing part of the problem is solved by exhaustive search.
In~\cite{MorozovHumphriesEtAl2019} high-fidelity nonlinear simulations are used to compare the design of the EMs and gear-ratios, and to search for the most advantageous drivetrain layout.
Finally, the authors of~\cite{VerbruggenSilvasEtAl2020} solve the joint e-powertrain design and control problem via derivative-free optimization for different architectures.
Overall, whilst all these methods in some form contribute to solve the problem of finding the optimal gear-ratios and control strategy, they are based on nonlinear, derivative-free and/or combinatorial optimization methods, or exhaustive search, resulting in high computation times without providing global optimality guarantees.

In conclusion, to the best of the authors' knowledge, a method to jointly optimize the design and control of an MGT for an EV in a computationally efficient manner is not available yet.

\subsubsection*{Statement of Contributions}
Against this backdrop, this paper presents a computationally efficient framework to jointly optimize the design and the control of the transmission of a central-drive EV on a given drive cycle.
To this end, we devise a convex yet high-fidelity EM model (see Fig.~\ref{fig:MotorMap}), and leverage it to compute the optimal CVT control trajectories and FGT design in closed form.
To tackle the complexity of jointly optimizing the gear-ratios and gear-shifting patterns of an MGT, we combine our closed-form solution for the FGT design with DP in an iterative fashion.
Finally, we validate and benchmark our methods with nonlinear simulations and mixed-integer quadratic programming (MIQP), respectively, and compare the achievable performance of an FGT-, CVT- and MGT-equipped EV in terms of energy efficiency. Thereby, the computational efficiency of our method enables us to also optimize the EM size via exhaustive search.

\subsubsection*{Organization}
The remainder of this paper is structured as follows: Section~\ref{sec:methodology} presents a model of a central-drive EV, including a convex EM model, and states the minimum-energy transmission design and control problem. We present solution algorithms to find the optimal gear designs and strategies for a CVT, FGT and MGT in Section~\ref{sec:solutionalgorithms}, whilst in Section~\ref{sec:results}, we offer a numerical case study comparing the three types of transmissions and present validation results.
Finally, Section~\ref{sec:conclusion} draws the conclusions.

\section{Methodology}
\label{sec:methodology}
This section derives a convex model of the EV shown in Fig.~\ref{fig:Veh_Model}, including the EM and the transmission, and formulates the energy-optimal transmission design and control problem for a CVT, an FGT and an MGT.
To this end, we leverage the quasistatic modeling approach from~\cite{GuzzellaSciarretta2007}.

\subsection{Vehicle and Transmission}
\label{sec:VehicleModel}
For a given discretized driving cycle consisting of a speed trajectory $v(t)$, acceleration trajectory $a(t)$ and road grade trajectory $\alpha (t)$, the required power at the wheels is
\par\nobreak\vspace{-5pt}
\begingroup
\allowdisplaybreaks
\begin{small}
\begin{equation}
    \begin{aligned}
P_{\mathrm{t}}(t)=& m_{\mathrm{v}} \cdot\left(c_{\mathrm{r}} \cdot g \cdot \cos (\alpha(t))+g \cdot \sin (\alpha(t))+a(t)\right) \cdot v(t) \\
&+\frac{1}{2} \cdot \rho \cdot c_{\mathrm{d}} \cdot A_{\mathrm{f}} \cdot v(t)^{3},
\end{aligned} 
\label{eq:dcycle}
\end{equation}
\end{small}%
\endgroup
where $m_\mathrm{v}$ is the total mass of the vehicle, $c_\mathrm{r}$ is the rolling friction coefficient, $g$ is the gravitational acceleration, $\rho_\mathrm{a}$ is the air density, $c_\mathrm{d}$ is the air drag coefficient and $A_\mathrm{f}$ is the frontal area of the vehicle.
Similarly, the required torque at the wheels is
\par\nobreak\vspace{-5pt}
\begingroup
\allowdisplaybreaks
\begin{small}
\begin{equation}
	\begin{aligned}
		T_{\mathrm{t}}(t)=& m_{\mathrm{v}} \cdot\left(c_{\mathrm{r}} \cdot g \cdot \cos (\alpha(t))+g \cdot \sin (\alpha(t))+a(t)\right).
	\end{aligned} 
	\label{eq:Tdcycle}
\end{equation}
\end{small}%
\endgroup
We assume the gearbox efficiency to have a different constant value for each transmission technology. This way, the EM power required to follow the drive cycle is
\par\nobreak\vspace{-5pt}
\begingroup
\allowdisplaybreaks
\begin{small}
\begin{equation}
    P_\mathrm{m}(t) = \left\{\begin{array}{ll}
P_\mathrm{t}(t) \cdot \eta_\mathrm{fgt}^{-\mathrm{sign}(P_\mathrm{t}(t))} & \text{ if FGT } \\ 
P_\mathrm{t}(t)\cdot \eta_\mathrm{mgt}^{-\mathrm{sign}(P_\mathrm{t}(t))} & \text{ if MGT } \\
P_\mathrm{t}(t) \cdot \eta_\mathrm{cvt}^{-\mathrm{sign}(P_\mathrm{t}(t))} & \text{ if CVT,}
\end{array}\right.
    \label{eq:Pmotor}
\end{equation}
\end{small}%
\endgroup
where $\eta_\mathrm{fgt}$, $\eta_\mathrm{mgt}$ and $\eta_\mathrm{cvt}$ are the efficiencies of the FGT, MGT and CVT, respectively.
Similarly, the EM torque at the wheels (i.e., without including the gear-ratio) is
\par\nobreak\vspace{-5pt}
\begingroup
\allowdisplaybreaks
\begin{small}
\begin{equation}
	T_\mathrm{m,w}(t) = \left\{\begin{array}{ll}
		T_\mathrm{t}(t) \cdot \eta_\mathrm{fgt}^{-\mathrm{sign}(T_\mathrm{t}(t))} & \text{ if FGT } \\ 
		T_\mathrm{t}(t) \cdot \eta_\mathrm{mgt}^{-\mathrm{sign}(T_\mathrm{t}(t))} & \text{ if MGT } \\
		T_\mathrm{t}(t) \cdot \eta_\mathrm{cvt}^{-\mathrm{sign}(T_\mathrm{t}(t))} & \text{ if CVT},
	\end{array}\right.
	\label{eq:Tmotorw}
\end{equation}
\end{small}%
\endgroup
Assuming the friction brakes to be used only when the EM is saturated at its minimum power or torque (in order to maximize regenerative braking), $P_\mathrm{m}(t)$ and $T_\mathrm{m,w}(t)$ can be known in advance and the operating point of the electric motor is solely determined by the transmission. Thereby, the rotational EM speed $\omega_\mathrm{m}(t)$ is related to the wheels' speed through the transmission ratio $\gamma(t)$ as
\par\nobreak\vspace{-5pt}
\begingroup
\allowdisplaybreaks
\begin{small}
\begin{equation}
    \omega_\mathrm{m}(t) = \gamma(t) \cdot \frac{v(t)}{r_\mathrm{w}} ,
    \label{eq:motorspeed}
\end{equation}
\end{small}%
\endgroup
where $r_\mathrm{w}$ is the wheels' radius. The transmission ratio (including the final drive) is subject to optimization and is constrained depending on the transmission technology as
\par\nobreak\vspace{-5pt}
\begingroup
\allowdisplaybreaks
\begin{small}
\begin{equation}
    \gamma(t)\left\{\begin{array}{ll}
=\gamma_{1} & \text{ if FGT } \\ 
\in \{ \gamma_1,... ,\gamma_{n_\mathrm{gears}} \} & \text{ if MGT } \\
\in\left[\gamma_{\mathrm{cvt},\min }, \gamma_{\mathrm{cvt},\max }\right] & \text{ if CVT,}
\end{array}\right.
\label{eq:transmission}
\end{equation}
\end{small}%
\endgroup
where $\gamma_1,...,\gamma_{n_\mathrm{gears}} $ are fixed gear-ratios of the FGT (for $n_\mathrm{gears}=1$) and the MGT and $n_\mathrm{gears}$ is the number of gears in the MGT, whilst $\gamma_\mathrm{min}$ and $\gamma_\mathrm{max}$ are the gear-ratio limits of the CVT which we consider as given.

For the sake of simplicity, as a performance requirement, we focus on the stand-still towing requirements on a slope of $\alpha_0$:
\par\nobreak\vspace{-5pt}
\begingroup
\allowdisplaybreaks
\begin{small}
\begin{equation}
    m_\mathrm{v}  \cdot g \cdot \sin{(\alpha_0)} \cdot r_\mathrm{w} \leq T_\mathrm{m,max} \cdot \left\{\begin{array}{ll}
\eta_\mathrm{fgt} \cdot \gamma_1 & \text{ if FGT } \\ 
\eta_\mathrm{mgt} \cdot \gamma_1 & \text{  if MGT} \\
\eta_\mathrm{cvt} \cdot \gamma_\mathrm{max} & \text{ if CVT}.
\end{array}\right.
\label{eq:SlopeStartConstraint}
\end{equation}
\end{small}%
\endgroup
However, our framework readily accommodates additional requirements which we leave to an extended version of this paper.

\subsection{Vehicle Mass}
The mass of the vehicle depends on the transmission used.
The total mass of the vehicle $m_\mathrm{v}$ is the sum of a base weight $m_0$, the weight of the gearbox $m_\mathrm{g}$ and the weight of the motor $m_\mathrm{m}$, yielding
\par\nobreak\vspace{-5pt}
\begingroup
\allowdisplaybreaks
\begin{small}
\begin{equation}
    m_\mathrm{v} = m_0 + m_\mathrm{g} + m_\mathrm{m} .
    \label{eq:vehiclemass}
\end{equation}
\end{small}%
\endgroup
Similarly to \cite{VerbruggenSalazarEtAl2019,BorsboomFahdzyanaEtAl2021}, the motor mass $m_\mathrm{m}$ is modeled linear in relation to the maximum motor power $P_\mathrm{m,max}$ as
\par\nobreak\vspace{-5pt}
\begingroup
\allowdisplaybreaks
\begin{small}
\begin{equation}
    m_\mathrm{m} = \rho_{\mathrm{m}}\cdot P_\mathrm{m,max} ,
\end{equation}
\end{small}%
\endgroup
where $\rho_\mathrm{m}$ represents the specific mass of the motor.
The mass of the gearbox is modeled linearly with the number of gears as
\par\nobreak\vspace{-5pt}
\begingroup
\allowdisplaybreaks
\begin{small}
\begin{equation}
    m_\mathrm{g} = \left\{\begin{array}{ll}
m_{\mathrm{g},0} + m_{\mathrm{g}} & \text{ if FGT } \\ 
m_{\mathrm{g},0} + m_{\mathrm{g}}\cdot n_\mathrm{gears}  & \text{ if MGT} \\
m_{\mathrm{cvt}} & \text{ if CVT,}
\end{array}\right.
\label{eq:transmissionmass}
\end{equation}
\end{small}%
\endgroup
where $m_{\mathrm{g},0}$ is the base mass for an FGT and an MGT, $m_{\mathrm{g}}$ is the added mass per gear, whilst $m_{\mathrm{cvt}}$ is the mass of a CVT.

\subsection{Motor Model}
\label{sec:MotorModel}
In general, the electric power $P_\mathrm{dc}$ is given by
\par\nobreak\vspace{-5pt}
\begingroup
\allowdisplaybreaks
\begin{small}
\begin{equation}
	P_\mathrm{dc}(t) = P_\mathrm{m}(t) + P_\mathrm{m,loss}(t),
	\label{eq:Ptotal}
\end{equation}
\end{small}%
\endgroup
where $P_\mathrm{m,loss}$ represents the EM losses.
While previous convex EM models, such as the quadratic model from~\cite{BorsboomFahdzyanaEtAl2021}, explicitly depend on both the EM speed and power, in this paper we devise a different approach based on the fact that the mechanical EM power $P_\mathrm{m}$ to be provided is known in advance.
Specifically, for each EM power value at time $t$, $P_\mathrm{m}$, we determine the power loss as a function of the EM speed $\omega_\mathrm{m}(t)$:
\par\nobreak\vspace{-5pt}
\begingroup
\allowdisplaybreaks
\begin{small}
\begin{equation}
\begin{split}
        P_\mathrm{m,loss}(t) = \frac{p_0(P_\mathrm{m}(t))}{\omega_\mathrm{m}} + p_1(P_\mathrm{m}(t)) + p_2(P_\mathrm{m}(t)) \cdot \omega_\mathrm{m} , \\
\end{split}
\label{eq:Plossmodel}
\end{equation}
\end{small}%
\endgroup
where the coefficients $p_0$, $p_1$ and $p_2$ are dependent on the motor power $P_\mathrm{m}(t)$. In practice, since the EM power is known in advance, the parameters can be provided for each time-step.
In order to ensure convexity for non-negative $\omega_\mathrm{m}$ values (which is the case in drive cycles), the parameters must satisfy
\par\nobreak\vspace{-5pt}
\begingroup
\allowdisplaybreaks
\begin{small}
\begin{equation}
    p_0 (P_\mathrm{m}) \geq 0\;\forall P_\mathrm{m}\neq 0, \quad p_0(0) =0,
    \label{eq:Constraintp0}
\end{equation}
\end{small}%
\endgroup
where the latter constraint makes sure that when the vehicle is standing still the losses do not diverge to infinity, and
\par\nobreak\vspace{-5pt}
\begingroup
\allowdisplaybreaks
\begin{small}
\begin{equation}
    p_2 (P_\mathrm{m}) \geq 0 \hspace{3pt} \forall P_\mathrm{m}.
    \label{eq:Constraintp2}
\end{equation}
\end{small}%
\endgroup
The motor operating point is bound by three factors: the maximum power, the maximum torque and the maximum motor speed. This leads to the following constraints: the constraint limiting the motor speed described by
\par\nobreak\vspace{-5pt}
\begingroup
\allowdisplaybreaks
\begin{small}
\begin{equation}
    \omega_\mathrm{m}(t) \in [0,\omega_\mathrm{m,max}] \hspace{1pt} ,
    \label{eq:maxmotorspeed} 
\end{equation}
\end{small}%
\endgroup
where $\omega_\mathrm{m,max}$ is the maximum motor speed; the constraint limiting the torque described by
\par\nobreak\vspace{-5pt}
\begingroup
\allowdisplaybreaks
\begin{small}
\begin{equation}
    T_\mathrm{m}(t) \in [-T_\mathrm{m,max},T_\mathrm{m,max}] ,
    \label{eq:maxtorque}
\end{equation}
\end{small}%
\endgroup
where $T_\mathrm{m,max}$ is the maximum torque. The maximum power constraint is given by
\par\nobreak\vspace{-5pt}
\begingroup
\allowdisplaybreaks
\begin{small}
\begin{equation}
    P_\mathrm{m}(t) \in [-P_\mathrm{m,max},P_\mathrm{m,max}],
    \label{eq:maxPower}
\end{equation}
\end{small}%
\endgroup
where $P_\mathrm{m,max}$ is constant.

\begin{figure}
    \centering
    \includegraphics[width = \columnwidth]{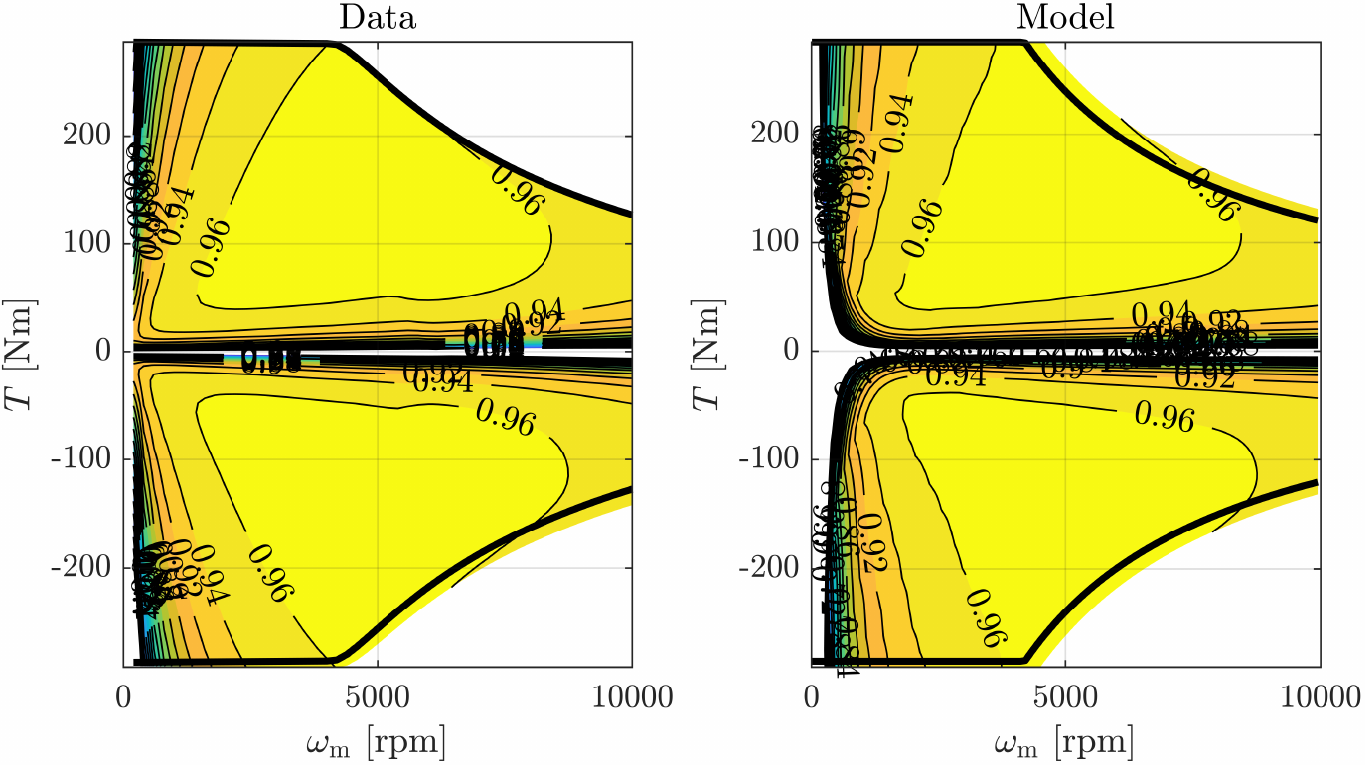}
    \caption{Motor efficiency maps resulting from data and from the model. The normalized RMSE w.r.t.\ the total electric power is 0.26\%.}
    \label{fig:MotorMap}
\end{figure}

The model described by \eqref{eq:Ptotal}--\eqref{eq:maxPower} is fitted to EM map data taken from MotorCAD~\cite{MotorCAD} shown on the left of Fig.~\ref{fig:MotorMap}. The resulting normalized root mean square error (RMSE) w.r.t.\ the total DC power is 0.26\%. The efficiency map resulting from the fitted model is shown to the right of Fig.~\ref{fig:MotorMap}.
All in all, we observe that our model not only provides an accurate fit, but also precisely captures the speed-dependence of the efficiency map.

\subsection{Problem Statement}
\label{sec:problemstatement}
Since the mechanical power of the EM $P_\mathrm{m}(t)$ is given, we minimize the energy usage of the EM by minimizing the sum of its losses $P_\mathrm{m,loss}(t)$. Specifically, we define our cost function as
\par\nobreak\vspace{-5pt}
\begingroup
\allowdisplaybreaks
\begin{small}
\begin{equation}\label{eq:totalEnergyLoss}
    J = \sum_{t = 0}^{T} P_\mathrm{m,loss}(t) \cdot  \Delta t + C_\mathrm{shift},
\end{equation}
\end{small}%
\endgroup
where the first term is the energy loss over the entire drive cycle and $C_\mathrm{shift}$ is a cost accounting for energy losses, drivability and components' wear when shifting gears. To this end, we introduce the time-dependent binary vector $b(t)\in\{0,1\}^{n_\mathrm{gears}}$, where $b_i (t) = 1$ if $ \gamma (t) = \gamma_i $ and 0 otherwise, and formally define $C_\mathrm{shift}$ as
\par\nobreak\vspace{-5pt}
\begingroup
\allowdisplaybreaks
\begin{small}
\begin{equation}
    C_\mathrm{shift} = \left\{\begin{array}{ll}
	0 & \text{ if FGT/CVT } \\ 
	c_\mathrm{shift} \cdot \frac{1}{2} \sum_{t = 1}^{T} \sum_{i=1}^{n_\mathrm{gears}}|\Delta b_i(t)| & \text{ if MGT},
\end{array}\right.
    \label{eq:ShiftCost}
\end{equation}
\end{small}%
\endgroup
where $\Delta b_i(t) = b_i(t) - b_i(t-1)$ and the constant $c_\mathrm{shift}$ captures the energy cost for a single gear-shift. We can now state the average energy-optimal transmission design and control problem as follows:




\begin{prob}[Minimum-energy Transmission Design and Control Problem]\label{prob:GeneralProblem}
	Given the electric vehicle architecture shown in Fig.~\ref{fig:Veh_Model} equipped with a CVT, FGT or MGT, the optimal gear-ratio(s) $\gamma_i^\star$ ($i=1$ for the FGT and $i=1,\dots,n_\mathrm{gears}$ for the MGT) and gear-trajectory $\gamma^\star(t)$ are the solution of
	\par\nobreak\vspace{-5pt}
	\begingroup
	\allowdisplaybreaks
	\begin{small}
	\begin{equation*}
		\begin{aligned}
			\min & \hspace{2pt} J\\
			\text{s.t. }&\eqref{eq:dcycle}-\eqref{eq:transmissionmass},\eqref{eq:Plossmodel},\eqref{eq:maxmotorspeed}-\eqref{eq:ShiftCost}.
		\end{aligned}
	\end{equation*}
	\end{small}%
	\endgroup
\end{prob}

\subsection{Discussion}
A few comments are in order. First, we assume the EM power as given. This assumption is acceptable for central-drive or symmetric in-wheels architectures without active torque vectoring, and enables to pre-compute the power demand on a drive cycle. Thereby, our method is readily applicable also in scenarios where only a predefined fraction of braking energy can be recuperated.
Second, we minimize the electric power provided to the EM, which is almost equivalent to the power provided by the battery, and approximate the shifting costs with a constant value, leaving a more careful analysis accounting for the battery efficiency and shifting dynamics to a journal extension of this paper.

\section{Solution Algorithms}
\label{sec:solutionalgorithms}
This section presents methods to efficiently solve Problem~\ref{prob:GeneralProblem} for a CVT, an FGT and an MGT.
\subsection{Energy-optimal Solution for the CVT}
\label{sec:CVT}
For the CVT, Problem~\ref{prob:GeneralProblem} consists of a pure optimal control problem.
Arguably, minimizing $J$ in~\eqref{eq:totalEnergyLoss} is equivalent to minimizing $P_\mathrm{m,loss}(t)$ at every time-instant. Rewriting the losses as a function of the gear-ratio $\gamma(t)$ we get
\par\nobreak\vspace{-5pt}
\begingroup
\allowdisplaybreaks
\begin{small}
	\begin{equation}
		P_\mathrm{m,loss}(t) = \frac{d_0(t)}{\gamma(t)} + d_1(t) + d_2(t) \cdot \gamma(t),
		\label{eq:Plossmodelconversion}
	\end{equation}
\end{small}%
\endgroup
where the coefficients $d_0(t)$, $d_1(t)$ and $d_2(t)$ are related to the coefficients from~\eqref{eq:Plossmodel} as $d_0(t) = \frac{p_0(P_\mathrm{m}(t))}{v(t)/r_\mathrm{w}}\geq0$, $d_1(t) = p_1(P_\mathrm{m}(t))$ and $d_2(t) = p_2(P_\mathrm{m}(t))\cdot \frac{v(t)}{r_\mathrm{w}}\geq0$. Crucially, since $P_\mathrm{m}(t)$ is known in advance, these coefficients' trajectory is also known in advance.

We rewrite constraints~\eqref{eq:maxmotorspeed}, \eqref{eq:maxtorque} and \eqref{eq:maxPower} for $\gamma$ as
\par\nobreak\vspace{-5pt}
\begingroup
\allowdisplaybreaks
\begin{small}
\begin{equation}
	\begin{aligned}
		\gamma(t) &\geq 0\\
		\gamma(t) &\geq \frac{|T_\mathrm{m,w}(t)|}{T_\mathrm{m,max}}\\
		\gamma(t) &\leq \,\omega_\mathrm{m,max}\cdot \frac{r_\mathrm{w}}{v(t)}.
		\label{eq:gammaconstraintsCVT}
	\end{aligned}
\end{equation}
\end{small}%
\endgroup
Therefore, the minimal and maximal values for $\gamma(t)$ are
\par\nobreak\vspace{-5pt}
\begingroup
\allowdisplaybreaks
\begin{small}
\begin{equation}
	\begin{aligned}
		\gamma_\mathrm{min}(t) &= \max\left(0,\frac{|T_\mathrm{m,w}(t)|}{T_\mathrm{m,max}} ,\gamma_{\mathrm{cvt},\min}\right)\\
		\gamma_\mathrm{max}(t) &= \mathrm{min}\left( \omega_\mathrm{m,max}\cdot \frac{r_\mathrm{w}}{v(t)}\right).
	\end{aligned}
	\label{eq:ymin}
\end{equation}
\end{small}%
\endgroup
Since we consider only the control of the CVT and constraint~\eqref{eq:SlopeStartConstraint} only affects transmission design, it is not taken into account here.

We first consider an unconstrained operation. Since $P_\mathrm{m,loss}$ is a convex function for $\gamma(t)\geq 0$, we set the derivative of \eqref{eq:Plossmodelconversion} equal to zero and we solve for $\gamma(t)$. This way, the optimal unconstrained gear-ratio is
\par\nobreak\vspace{-5pt}
\begingroup
\allowdisplaybreaks
\begin{small}
\begin{equation}\label{eq:ucoptimumCVT}
        \gamma_\mathrm{uc}^\star(t)  = \sqrt{\frac{d_0(t)}{d_2(t)}}.
\end{equation}
\end{small}%
\endgroup
Since~\eqref{eq:Plossmodelconversion} is convex for non-negative arguments and is a function of a single optimization variable, its global optimum $\gamma^\star(t)$ corresponds to the feasible value that is closest to its unconstrained minimizer $\gamma_\mathrm{uc}^\star(t)$.
Therefore, the optimal gear-ratio $\gamma^\star$ is found as
\par\nobreak\vspace{-5pt}
\begingroup
\allowdisplaybreaks
\begin{small}
 \begin{equation}\label{eq:optimumCVT}
     \gamma^\star(t) = \mathrm{min}(\gamma_\mathrm{max}(t),\mathrm{max}(\gamma_\mathrm{min}(t),\gamma_\mathrm{uc}^\star(t))).
 \end{equation}
 \end{small}%
\endgroup
This way, the global optimum of Problem~\ref{prob:GeneralProblem} can be efficiently computed by simple matrix operations.



\subsection{Energy-optimal Solution for the FGT}
\label{sec:FGT}
In order to solve Problem~\ref{prob:GeneralProblem} for the FGT, we use a similar approach as in Section~\ref{sec:CVT} above. In this case, problem~\ref{prob:GeneralProblem} is reduced to a pure optimal design problem, since the gear-ratio $\gamma$ is constant throughout the drive cycle.
Also in this case, we do not have gear-shift costs. Hence the objective $J$ in~\eqref{eq:totalEnergyLoss} consists of energy losses only, which we rewrite as
\par\nobreak\vspace{-5pt}
\begingroup
\allowdisplaybreaks
\begin{small}
\begin{equation}
\begin{split}
        \sum_{t = 0}^{T} P_\mathrm{m,loss}(t) \cdot  \Delta t  = \frac{e_0}{\gamma} + e_1 + e_2 \cdot \gamma  ,
\end{split}
\label{eq:PlossFGT}
\end{equation}
\end{small}%
\endgroup
where the new parameters ${e_0 = \sum_{t = 0}^{T} d_0(t) \cdot\Delta t\geq0}$, ${ e_1 = \sum_{t = 0}^{T} d_1(t) \cdot\Delta t }$ and ${e_2 = \sum_{t = 0}^{T} d_2(t) \cdot\Delta t \geq0}$ are known in advance.

We rewrite constraints~\eqref{eq:gammaconstraintsCVT}, including~\eqref{eq:SlopeStartConstraint} as
\par\nobreak\vspace{-5pt}
\begingroup
\allowdisplaybreaks
\begin{small}
\begin{equation}
	\begin{aligned}
		\gamma &\geq 0\\
		\gamma &\geq \frac{|T_\mathrm{m,w}(t)|}{T_\mathrm{m,max}}\\
		\gamma &\geq \frac{m_\mathrm{v}  \cdot g \cdot \sin{(\alpha_0)} \cdot r_\mathrm{w}} {T_\mathrm{m,max} \cdot \eta_\mathrm{fgt}}\\
		\gamma &\leq \,\omega_\mathrm{m,max}\cdot \frac{r_\mathrm{w}}{v(t)}
	\end{aligned}
	\label{eq:gammaconstraintPmaxFGT}
\end{equation}
\end{small}%
\endgroup
Therefore, the minimum and maximum for $\gamma$ are
\par\nobreak\vspace{-5pt}
\begingroup
\allowdisplaybreaks
\begin{small}
\begin{equation}
	\begin{aligned}
		\gamma_\mathrm{min} &= \max_t \left( 0,\frac{|T_\mathrm{m,w}(t)|}{T_\mathrm{m,max}},  \frac{m_\mathrm{v}  \cdot g \cdot \sin{(\alpha_0)} \cdot r_\mathrm{w}} {T_\mathrm{m,max} \cdot \eta_\mathrm{fgt}} \right)\\
		\gamma_\mathrm{max} &= \min_t\left(\omega_\mathrm{m,max}\cdot \frac{r_\mathrm{w}}{v(t)}\right).
		\label{eq:minmaxFGT}
	\end{aligned}
\end{equation}
\end{small}%
\endgroup
Now we can use a similar approach as in Section~\ref{sec:CVT} to compute the energy-optimal gear ratio.
Since~\eqref{eq:PlossFGT} is convex for $\gamma\geq 0$, we set its derivative equal to zero to compute its unconstrained minimum as
\par\nobreak\vspace{-5pt}
\begingroup
\allowdisplaybreaks
\begin{small}
\begin{equation}
    \begin{split}
        \gamma_\mathrm{uc}^\star  = \sqrt{\frac{e_0}{e_2}} .
    \end{split}
\end{equation}
\end{small}%
\endgroup
Since~\eqref{eq:PlossFGT} is convex for non-negative arguments, the constrained optimum is the point that is closest to the unconstrained optimum whilst still satisfying the constraints, i.e.,
\par\nobreak\vspace{-5pt}
\begingroup
\allowdisplaybreaks
\begin{small}
 \begin{equation}\label{eq:optimumFGT}
	\gamma^\star = \mathrm{min}(\gamma_\mathrm{max},\mathrm{max}(\gamma_\mathrm{min},\gamma_\mathrm{uc}^\star)) .
\end{equation}
\end{small}%
\endgroup
Also for the FGT, the global optimum~\eqref{eq:optimumFGT} can be computed entirely by effective matrix operations.

\subsection{Energy-optimal Solution for the MGT}
\label{sec:MGT}

For the MGT, solving Problem~\ref{prob:GeneralProblem} is more intricately involved than for the CVT and FGT. This is because the problem is combinatorial: If the gear ratios are changed, the optimal gear-shift trajectory changes as well and vice versa. This might be solved using mixed-integer programming resulting in high computation times and potential convergence issues---as shown in the
\ifextendedversion
Appendix~\ref{sec:MIQP}.
\else
extended version of this paper~\cite{HurkSalazar2021}.
\fi
To overcome this limitation, we decouple the problem into its design and its control aspects.
Specifically, we first devise a method to compute the optimal gear-shifting trajectory for given gear-ratios, and then formulate an approach inspired by Section~\ref{sec:FGT} above to compute the optimal gear-ratios for a given gear-shifting trajectory.
Finally, we iterate between these two methods until convergence to find an optimal design and control solution of Problem~\ref{prob:GeneralProblem} for the MGT.

\subsubsection{Optimal Gear-shifting Trajectory}
\label{sec:MGTgeartrajectory}
We consider the set of gear ratios $\gamma_i$ to be given, for which we have to solve problem~\ref{prob:GeneralProblem} to find the optimal gear trajectory.
This way, the objective $J$ is solely influenced by the gear-shifting trajectory defined by the binary variable $b_i(t)$ introduced in Section~\ref{sec:problemstatement}.
If we neglect gear-shifting costs setting $c_\mathrm{shift}=0$, we can minimize the $J$ by determining the resulting $P_\mathrm{m,loss}(t)$ for each gear $\gamma_i$ and choose $b_i^\star(t) = 1$ if the corresponding gear $\gamma_i$ results in the lowest feasible power loss at time-step $t$. This approach entails choosing the minimum element in $T+1$ vectors of size $n_\mathrm{gears}$, which is parallelizable and can be solved extremely efficiently.
If gear-shifting costs are included ($c_\mathrm{shift}>0$), Problem~\ref{prob:GeneralProblem} can be rapidly solved with dynamic programming due to the presence of only one state and input variable constrained to sets with cardinality $n_\mathrm{gears}$.

\subsubsection{Optimal Gear-ratios}
\label{sec:MGTGearRatios}
We consider the gear-shifting trajectory defined by the binary variable $b_i(t)$ to be given.
Since the gear trajectory is given, $C_\mathrm{shift}$ is constant and, therefore, minimizing $J$ corresponds to minimizing the energy losses~\eqref{eq:Plossmodelconversion}, which we rewrite as
\par\nobreak\vspace{-5pt}
\begingroup
\allowdisplaybreaks
\begin{small}
\begin{equation}
\begin{split}
        \sum_{t=0}^T P_\mathrm{m,loss}(t)\cdot\Delta t = \sum_{i=1}^{n_\mathrm{gears}} \frac{f_{0,i}}{\gamma_i} + f_{1,i} + f_{2,i} \cdot \gamma_i , 
\end{split}
\label{eq:PlossMGT}
\end{equation}
\end{small}%
\endgroup
where the constant coefficients
\par\nobreak\vspace{-5pt}
\begingroup
\allowdisplaybreaks
\begin{small}
\begin{equation*}
	\begin{aligned}
	f_{0,i} &= \sum_{t=0}^T d_0(P_\mathrm{m}(t)) \cdot b_i(t)\geq 0\\
	f_{1,i} &= \sum_{t=0}^T d_1(P_\mathrm{m}(t)) \cdot b_i(t)\\
	f_{2,i} &= \sum_{t=0}^T d_2(P_\mathrm{m}(t))\cdot b_i(t)\geq 0
\end{aligned}
\end{equation*}\\
\end{small}%
\endgroup
can be computed in advance.

We adapt the bounds for each $\gamma_i$ from~\eqref{eq:minmaxFGT}, considering that the constraints must hold only for the selected gear: 
\par\nobreak\vspace{-5pt}
\begingroup
\allowdisplaybreaks
\begin{small}
\begin{equation}
		\begin{aligned}
			\gamma_{\mathrm{min},i} &= \max_{t:b_{i}(t) = 1}\left(0,\frac{|T_\mathrm{m,w}(t)|}{T_\mathrm{m,max}}\cdot \right)\;\forall i=\{2,...,n_\mathrm{gears}\}\\
			\gamma_{\mathrm{max},i} &= \min_{t:b_{i}(t) = 1}\left(\omega_\mathrm{m,max}\cdot \frac{r_\mathrm{w}}{v(t)}\right) \forall i=\{1,...,n_\mathrm{gears}\}.
		\end{aligned}
	\label{eq:minmaxMGT}
\end{equation}
\end{small}%
\endgroup
Hereby, the towing constraint~\eqref{eq:SlopeStartConstraint} only holds in first gear:
\par\nobreak\vspace{-5pt}
\begingroup
\allowdisplaybreaks
\begin{small}
\begin{equation}
		\gamma_{\mathrm{min},1}  = \max_{t:b_{1}(t) = 1}\left(0, \frac{|T_\mathrm{m,w}(t)|}{T_\mathrm{m,max}} , \frac{m_\mathrm{v}  \cdot g \cdot \sin{(\alpha_0)} \cdot r_\mathrm{w}} {T_\mathrm{m,max} \cdot \eta_\mathrm{mgt}}\right).
		\label{eq:minMGT1}
\end{equation}
\end{small}%
\endgroup
The energy-optimal gear-ratios can be found in a similar fashion to Sections~\ref{sec:CVT} and \ref{sec:FGT}. Since \eqref{eq:PlossMGT} is convex for $\gamma_i\geq 0$, we set its derivative equal to zero to compute the unconstrained optimum as
\par\nobreak\vspace{-5pt}
\begingroup
\allowdisplaybreaks
\begin{small}
\begin{equation}
    \begin{split}
        \gamma_{\mathrm{uc},i}^\star  = \sqrt{\frac{f_{0,i}}{f_{2,i}}}  \quad\forall i=\{1,...,n_\mathrm{gears}\}. \\
    \end{split}
\end{equation}
\end{small}%
\endgroup
Hereby, since~\eqref{eq:PlossMGT} corresponds to the sum of $n_\mathrm{gears}$-times one-dimensional problems that are convex for non-negative arguments, its constrained minimum is the point that is closest to the unconstrained optimum whilst still satisfying the constraints
 \par\nobreak\vspace{-5pt}
\begingroup
\allowdisplaybreaks
\begin{small}
 \begin{equation}
     \gamma_i^\star = \mathrm{min}(\gamma_{\mathrm{max},i} ,\mathrm{max}(\gamma_{\mathrm{min},i},\gamma_{\mathrm{uc},i}^\star)) \quad\forall i=\{1,...,n_\mathrm{gears}\}.
     \label{eq:constrainedoptimumMGT}
 \end{equation}
 \end{small}%
\endgroup
 
\subsubsection{Finding the Optimal Gear Strategy and Design}
We finally solve Problem~\ref{prob:GeneralProblem} for the MGT iterating on the methods presented in Section~\ref{sec:MGTgeartrajectory} and~\ref{sec:MGTGearRatios}, as shown in Algorithm~\ref{alg:Iterative}.
Thereby, we terminate the iteration when the relative difference in the cost function $J$ is smaller than a tolerance~$\epsilon$.

\begin{algorithm}[t!]
	\caption{Iterative Algorithm}
	\label{alg:Iterative}
	\begin{algorithmic}
	    \State $\gamma_i^\star    = \gamma_{i,\mathrm{start}} $ 
        \While{ $ \| J - J_\mathrm{prev} \| \geq \epsilon\cdot\|J\|$}
        \State $J_\mathrm{prev} = J$
		\State Compute $b_i^\star(t)$ for $\gamma_i^\star$ as in Section~\ref{sec:MGTgeartrajectory}
		\State Compute $\gamma_i^\star$ for $b_i^\star(t)$ as in~\eqref{eq:constrainedoptimumMGT}
		\State Compute $J$ for $b_i^\star(t)$ and $\gamma_i^\star$ as in~\eqref{eq:totalEnergyLoss}
		\EndWhile
	\end{algorithmic}
\end{algorithm}

\subsection{Note on Optimality}
Whilst the methods devised to solve Problem~\ref{prob:GeneralProblem} for the FGT and CVT owe global optimality guarantees to the convex problem's structure, the iterative approach presented in Section~\ref{sec:MGT} to solve the combinatorial Problem~\ref{prob:GeneralProblem} for an MGT sacrifices global optimality for computational efficiency.
However, the optimality benchmark provided in Section~\ref{sec:validation} is very promising.

\section{Results}
\label{sec:results}
In this section we showcase our algorithms with a numerical case study for a compact EV, and validate them via mixed-integer convex programming and nonlinear simulations. Specifically, we consider a compact family car completing the Class 3 Worldwide harmonized Light-duty vehicles Test Cycle (WLTC).
The vehicle parameters are provided in
\ifextendedversion
 Table~\ref{tab:vehicleparams} in Appendix~\ref{sec:ResultTables}.
 \else
 the extended version of this paper~\cite{HurkSalazar2021}.
\fi

The methods devised to solve Problem~\ref{prob:GeneralProblem} for a CVT, an FGT and an MGT, given the gear-shifting trajectory, do not require particular numerical solvers and can be efficiently solved by matrix operations only, whilst finding the optimal gear-shifting trajectory of an MGT with DP takes approximately \unit[7]{ms}.
When solving the joint design and control Problem~\ref{prob:GeneralProblem} for an MGT, our iterative Algorithm~\ref{alg:Iterative} converges in about~\unit[180]{ms} and 15 iterations on average.
Because of this very low solving time, we can jointly optimize the EM sizing by bruteforcing it.
To this end, we scale the EM linearly with its maximum power similarly to~\cite{VerbruggenSalazarEtAl2019} and as shown in
\ifextendedversion
Appendix~\ref{sec:motorsizing}.
\else
the extended version of this paper~\cite{HurkSalazar2021}.
\fi
However, given the bruteforce approach, our Algorithm~\ref{alg:Iterative} could be used just as effectively with a finite number of EM designs, for instance pre-computed with a finite-element software such as MotorCAD~\cite{MotorCAD}.
We run our algorithm for 100 different EM sizes. Including bruteforcing the EM size, the solution of the joint EM and transmission design and control problem is computed in a few seconds, with the longest run time of \unit[9]{s} for the 5-speed MGT.

\subsection{Compact Car Case Study}
\label{sec:numericalresults}

\begin{figure}
    \centering
    \includegraphics[width = \linewidth]{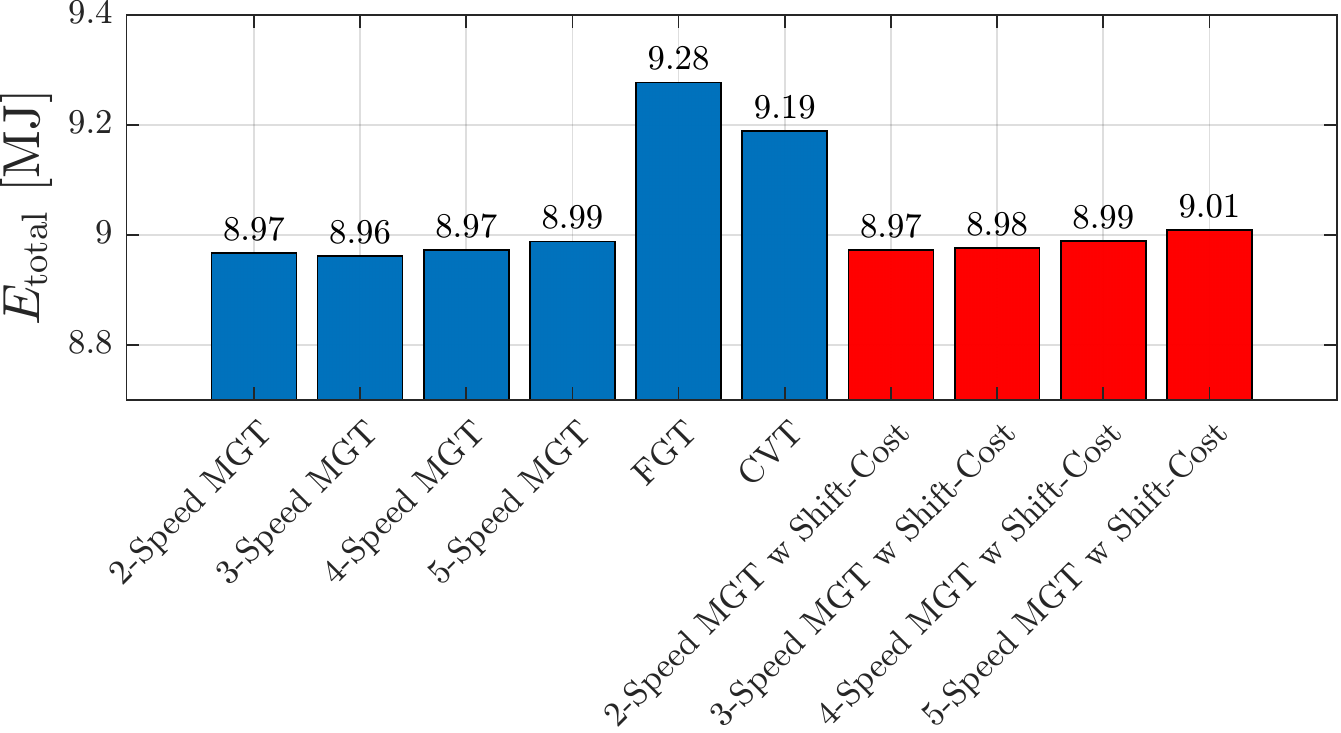}
    \caption{Optimal total energy usage for the FGT, CVT and MGTs.}
    \label{fig:ResultChart}
\end{figure}
\begin{figure}
	\centering
	\includegraphics[width = \columnwidth]{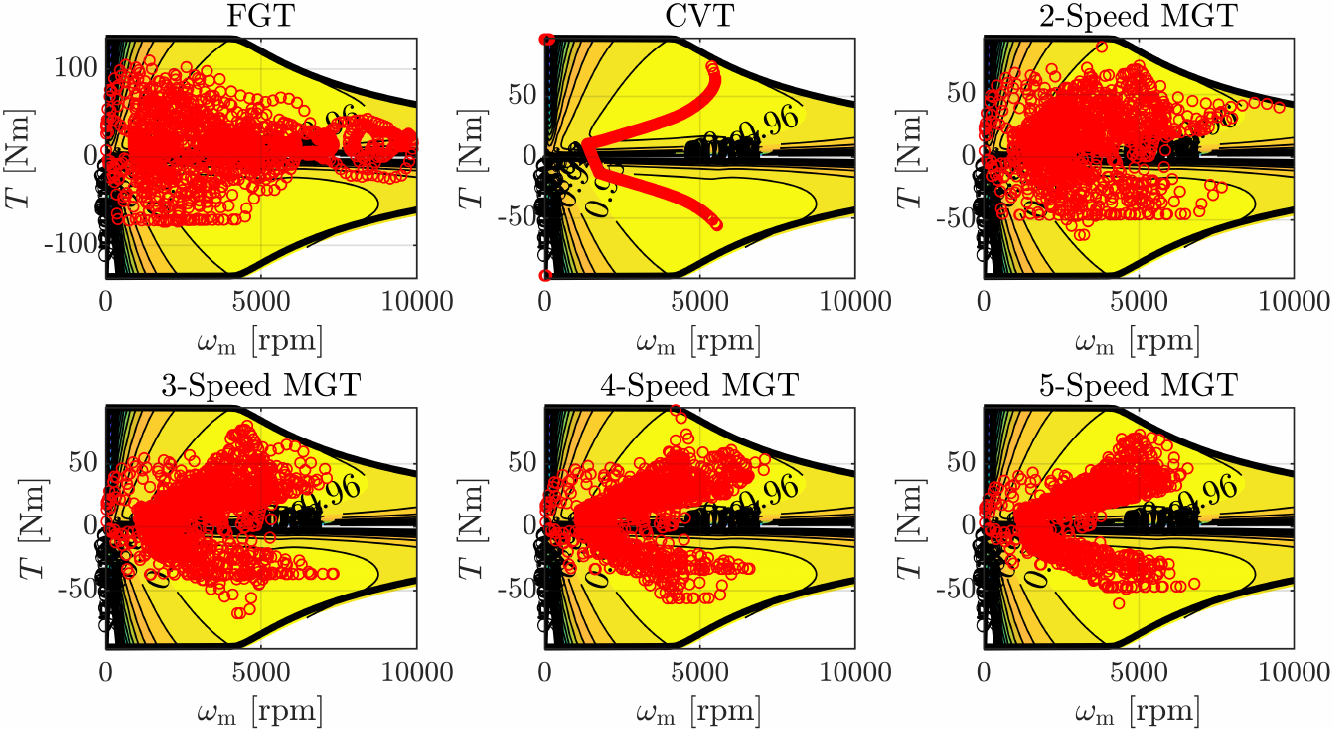}
	\caption{Resulting operating points for the FGT, CVT and MGTs without gear-shifting costs plotted over the scaled original EM map.}
	\label{fig:OperatingPoints}
\end{figure}

Fig.~\ref{fig:ResultChart} summarizes the results obtained for a 2-speed to 5-speed MGT (with and without gear-shifting costs), an FGT and a CVT, whereby the EM size is jointly optimized. The resulting EM operating points are shown in Fig.~\ref{fig:OperatingPoints}.
In addition, the resulting power loss, vehicle mass and EM size can be found in
\ifextendedversion
Table~\ref{tab:Results} in Appendix~\ref{sec:ResultTables}.
\else
the extended version of this paper~\cite{HurkSalazar2021}.
\fi

We observe the FGT yielding the worst performance. The main reason for this is that, owing to the towing constraint~\eqref{eq:SlopeStartConstraint}, the EM needs to be able to deliver a high torque and therefore has a larger size compared to the other transmissions. This effect, combined with the FGT inability of choosing the EM speed in turn results in a lower number of operating points lying in the EM high-efficiency region.
Interestingly, the CVT is also outperformed by every MGT, as the additional benefits of choosing any gear-ratio do not compensate for its higher weight and lower transmission efficiency.
For the MGTs, we observe the operating points converging to the areas of high efficiency as the number of gears is increased.

The benefits of adding gear ratios reach a point of diminishing returns between a 3-speed and a 4-speed MGT.
Whilst the energy losses decrease as the number of gears is increased, the increase in required mechanical energy due to the increased transmission mass causes an overall higher energy consumption.

Finally, also when including a small gear-shifting cost, the MGTs still outperform the CVT and the FGT. As a result of the decreased benefit of gear changes, the best performing transmission changes from the 3-speed MGT to the 2-speed MGT.
Further figures of the EM speed and the gear-ratio trajectories over the drive cycle can be found in
\ifextendedversion
Appendix~\ref{sec:speedandgear}.
\else
the extended version of this paper~\cite{HurkSalazar2021}.
\fi

\subsection{Validation}
\label{sec:validation}
To benchmark the optimality of our methods, we first compute the globally optimal solution for a slightly less accurate quadratic version of the EM model which can be solved to global optimality with mixed-integer quadratic programming (MIQP)---the details on the implementation can be found in
\ifextendedversion
Appendix~\ref{sec:MIQP}.
\else
extended version of this paper~\cite{HurkSalazar2021}.
\fi
Second, we apply the proposed methods to the same quadratic motor model and compare our results to the global optima computed via MIQP.
For the FGT and the CVT, the results obtained with quadratic programming and the results obtained with the proposed methods applied on the quadratic model are identical.
Solving the MIQP for a 2-speed MGT with Gurobi~\cite{GurobiOptimization2021} took \unit[15]{min} to converge for a fixed EM size, whilst no convergence was achieved for MGTs with 3 gear-ratios or more.
The gear ratios obtained with MIQP for a 2-speed MGT are $\gamma_1^\star = 6.14$ and $\gamma_2^\star = 14.40$, whilst our Algorithm~\ref{alg:Iterative} converged to $\gamma_1^\star = 6.09$ and $\gamma_2^\star = 14.44$.
The resulting total energy consumption differs less than 0.03\%: a difference which can be ascribed to numerical tolerances, indicating a good agreement between our solution and the global optimum.

\begin{figure}
	\centering
	\includegraphics[width = \linewidth]{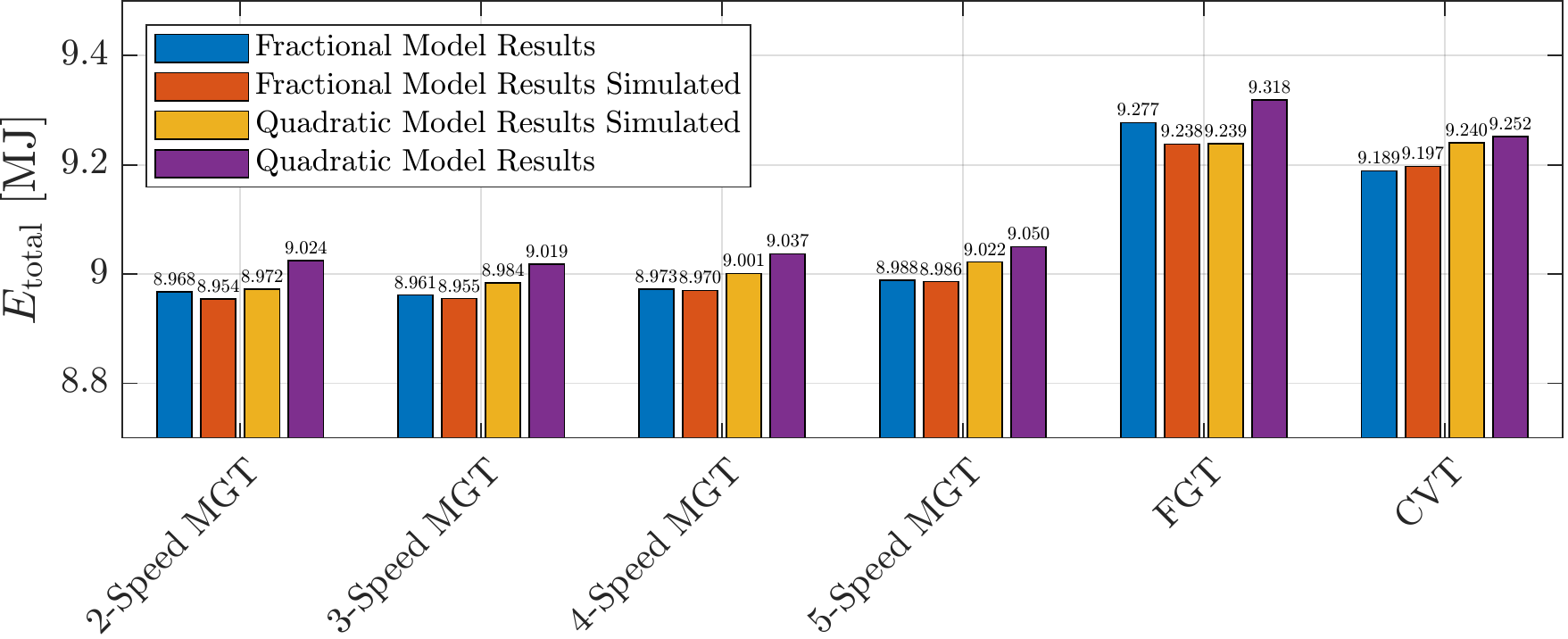}
	\caption{Comparison of the results for the FGT, MGT and CVT for the fractional model and the quadratic model, and the corresponding simulations.}
	\label{fig:SimResultChart}
\end{figure}
Finally, in order to validate the results stemming from different EM models, the results obtained with the fractional EM model~\eqref{eq:Plossmodel} and the quadratic model are simulated using the original nonlinear motor data. This final step on the one hand assesses the performance that would be achieved in practice, and on the other hand validates the accuracy of our models.
Fig.~\ref{fig:SimResultChart} summarizes the results obtained with our methods and their performance in simulation, showing that our proposed fractional EM model~\eqref{eq:Plossmodel} can predict the energy consumption of the EM very accurately, outperforming the quadratic model for all the transmission technologies.
Additional results including the optimal EM size and vehicle mass are provided in
\ifextendedversion
Table~\ref{tab:Results} in Appendix~\ref{sec:ResultTables},
\else
extended version of this paper~\cite{HurkSalazar2021},
\fi
where we observe that both the proposed EM model and the quadratic model always result in the same EM size.

\section{Conclusions}
\label{sec:conclusion}
This paper explored the possibility of jointly optimizing the design and control of an electric vehicle's transmission combining convex analysis and dynamic programming. Our algorithms resulted in a computational time in the order of \unit[100]{ms}, enabling us to also optimize the electric motor size via exhaustive search---we could compare 100 different sizes in less than \unit[10]{s}.
Our modeling approach was validated with nonlinear simulations showing an overall high accuracy, whilst the optimal solution found by our algorithm was in line with the global optimum computed via mixed-integer quadratic programming.
On the design side, our results showed that for the electric vehicles under consideration, a 2-speed transmission appears to be the most efficient solution when compared to other technologies, striking the best trade-off in terms of components' mass and control flexibility.

This work opens the field for several extensions.
First, the relatively high fidelity of the proposed convex models could be leveraged for more general powertrain control and optimization problems~\cite{KorziliusBorsboomEtAl2021}.
Second, the methods could be extended to account for thermal dynamics~\cite{LocatelloKondaEtAl2020}.
Finally, the computational efficiency of our approach could be leveraged both in more comprehensive design studies (e.g., including the battery dynamics) and real-time control applications.


\section*{Acknowledgment}

We would like to thank Mr.\ Olaf Korzilius, Dr.\ Steven Wilkins and Dr.\ Pascal Etman for the fruitful discussions as well as Dr.\ Ilse New for proofreading this paper.



%


\bibliographystyle{IEEEtran}
\bibliography{../../../Bibliography/main,../../../Bibliography/SML_papers}

\ifextendedversion

\newpage
\begin{appendices}
\section{Electric Motor Sizing}
\label{sec:motorsizing}
Similar to \cite{VerbruggenSalazarEtAl2019}, we scale the EM linearly with its maximum power. The motor will be scaled along a scaling parameter $s$ where 
\par\nobreak\vspace{-5pt}
\begingroup
\allowdisplaybreaks
\begin{small}
\begin{equation}
    s = \frac{P_\mathrm{m,max}}{\overline{P}_\mathrm{m,max}} .
\end{equation}
\end{small}%
\endgroup
The coefficients are then scaled as 
\par\nobreak\vspace{-5pt}
\begingroup
\allowdisplaybreaks
\begin{small}
\begin{equation}
    \begin{split}
    p_i = s \cdot \overline{p}_i \hspace{5pt} \forall i \in {0,1,2} \\
    c_{\mathrm{m},i} = s \cdot \overline{c}_{\mathrm{m},i} \hspace{5pt}  \forall i \in {0,1} .
    \end{split}
\end{equation}
\end{small}%
\endgroup

\section{Energy-optimal Solution of Problem~\ref{prob:GeneralProblem} via Mixed-integer Quadratic Programming}
\label{sec:MIQP}
The iterative Algorithm~\ref{alg:Iterative} we devised to jointly optimize the design and operation of an MGT does not provide global optimality guarantees.
However, we can perform a numerical benchmark to check the proximity of our solution to the global optimum obtained with mixed-integer convex programming.
However, given the mathematical structure of our problem, to the best of the authors' knowledge there are no off-the-shelf mixed-integer convex programming algorithms available that could solve it to global optimality.
Therefore, we use a slightly less accurate quadratic version of the EM model and solve Problem~\ref{prob:GeneralProblem} both with our Algorithm~\ref{alg:Iterative} and mixed-integer quadratic programming (MIQP).
Specifically, we model the EM as
\par\nobreak\vspace{-5pt}
\begingroup
\allowdisplaybreaks
\begin{small}
\begin{equation}
\begin{split}
        P_\mathrm{m,loss}(t) =  p_1(P_\mathrm{m}(t)) + p_2(P_\mathrm{m}(t)) \cdot \omega_\mathrm{m} + p_3(P_\mathrm{m}(t)) \cdot \omega_\mathrm{m}^2 ,
\end{split}
\end{equation}
\end{small}%
\endgroup
where $p_1(P_\mathrm{m}(t))$, $p_2(P_\mathrm{m}(t))$ and $p_3(P_\mathrm{m}(t))\geq 0$ are determined as in Section~\ref{prob:GeneralProblem}, and then frame Problem~\ref{prob:GeneralProblem} as a MIQP using the big-$M$ method~\cite{RichardsHow2005}.
First, we relax the losses as
\par\nobreak\vspace{-5pt}
\begingroup
\allowdisplaybreaks
\begin{small}
	\begin{equation}
		\begin{split}
			P_\mathrm{m,loss}(t) \geq  p_1(P_\mathrm{m}(t)) + p_2(P_\mathrm{m}(t)) \cdot \omega_\mathrm{m} + p_3(P_\mathrm{m}(t)) \cdot \omega_\mathrm{m}^2 ,
		\end{split}
		\label{eq:QuadraticPlossmodel}
	\end{equation}
\end{small}%
\endgroup
This relaxation is lossless and will hold with equality if the total energy is minimized.
The EM speed $\omega_\mathrm{m}(t)$ is determined by
\par\nobreak\vspace{-5pt}
\begingroup
\allowdisplaybreaks
\begin{small}
\begin{equation}
	\begin{aligned}
    \omega_m(t) &\geq b_i(t) \cdot  \gamma_i \cdot \frac{v(t)}{r_w} - b_i(t)\cdot M\\
    \omega_m(t) &\leq b_i(t) \cdot  \gamma_i \cdot \frac{v(t)}{r_w} + b_i(t)\cdot M,
	\end{aligned}
    \label{eq:bigM_motorspeed2}
\end{equation}
\end{small}%
\endgroup
and we ensure that one gear is selected with
\par\nobreak\vspace{-5pt}
\begingroup
\allowdisplaybreaks
\begin{small}
\begin{equation}
    \sum_i^{n_\mathrm{gears}} b_i(t) = 1.
    \label{eq:BigM_sumbi}
\end{equation}
\end{small}%
\endgroup
Finally, we formulate the objective function as the sum of the EM losses as
\par\nobreak\vspace{-5pt}
\begingroup
\allowdisplaybreaks
\begin{small}
\begin{equation}
J = \sum_{t=0}^T P_\mathrm{m,loss}(t)\cdot  \Delta t + c_\mathrm{shift} \cdot \frac{1}{2} \sum_{t = 1}^{T} \sum_{i=1}^{n_\mathrm{gears}}|\Delta b_i(t)|,
\label{eq:ObjectiveFunction_MIQP}
\end{equation}
\end{small}%
\endgroup
where $\Delta b_i(t) = b_i(t) - b_i(t-1)$, and frame Problem~\ref{prob:GeneralProblem} as an MIQP as follows.
\begin{prob}[Mixed-Integer Quadratic Problem]\label{prob:MIQP}
	Given the electric vehicle architecture of Fig.~\ref{fig:Veh_Model} equipped with an MGT, the optimal gear-ratios $\gamma_i^\star$ and gear-trajectory $\gamma^\star(t)$ are the solution of
	\par\nobreak\vspace{-5pt}
\begingroup
\allowdisplaybreaks
\begin{small}
	\begin{equation*}
		\begin{aligned}
			\min & \hspace{2pt} J\\
			\text{s.t. }& \eqref{eq:dcycle}-\eqref{eq:transmissionmass},\eqref{eq:maxmotorspeed}-\eqref{eq:maxPower}, \eqref{eq:QuadraticPlossmodel}-\eqref{eq:ObjectiveFunction_MIQP}.
		\end{aligned}
	\end{equation*}
	\end{small}%
\endgroup
\end{prob}

\section{Vehicle Parameters and Optimization Results}
\label{sec:ResultTables}
\begin{table}[H]
\caption{Vehicle Parameters.}
\begin{tabular}{ p{0.3\linewidth} p{0.15\linewidth}  p{0.15\linewidth}  p{0.15\linewidth} }
\textbf{Parameter}                     & \textbf{Symbol} & \textbf{Value} & \textbf{Unit} \\ \hline \vspace{0.5pt}
Vehicle base mass                      & $m_0$                &   1450            &     kg          \\
CVT mass                               & $m_\mathrm{cvt}$       &   80             &     kg          \\
FGT and MGT base mass                  & $m_{\mathrm{g},0}$       &   50             &     kg          \\
added mass per gear                    & $m_{\mathrm{gw}}$       &   5             &     kg/gear          \\
CVT efficiency                         & $\eta_\mathrm{cvt}$     &   0.96             &     -          \\
FGT efficiency                         & $\eta_\mathrm{fgt}$    &   0.98             &     -          \\
MGT efficiency                         & $\eta_\mathrm{mgt}$    &   0.98             &     -          \\
Wheel radius                           &    $r_\mathrm{w}$             &    0.316            &       m        \\
Air drag coefficient                   &    $c_\mathrm{d}$             & 0.29           &      -         \\
Frontal area                           &     $A_\mathrm{f}$            & 0.725           &      m$^2$         \\
Air density                            &      $\rho$          &        1.25        &      kg/$m^2$         \\
Rolling resistance coefficient         &       $c_\mathrm{r}$          &      0.02          &      -         \\
Gravitational constant                 &        $g$           &       9.81         &     m/$s^2$        \\
Minimal Slope                          &    $\alpha_0$         &       25         &     \degree    \\
Gearshift Cost                          &    $c_\mathrm{shift}$         &       300          &     J    \\ 
Threshold Parameter                     &    $\epsilon$         &       0.0001          &     \%    \\ 
\\ \\

\end{tabular}
\label{tab:vehicleparams}
\end{table}

\begin{table}[H]
\caption{Optimization Results.}
\begin{tabular}{ p{0.20\linewidth} p{0.15\linewidth} p{0.15\linewidth}  p{0.14\linewidth}  p{0.14\linewidth}  }
\textbf{Transmission}    & \textbf{Total} \newline \textbf{Energy} & \textbf{Energy Loss} & \textbf{Vehicle Mass} & \textbf{Max Power} \\ \hline \\
FGT                      & \unit[9.277]{MJ}         &       \unit[804.5]{kJ}            &   \unit[1563]{kg}                &     \unit[64]{kW}          \\
2-Speed MGT                 & \unit[8.968]{MJ}      &       \unit[548.9]{kJ}            &   \unit[1551]{kg}                &     \unit[45]{kW}          \\
3-Speed MGT                 & \unit[8.962]{MJ}      &       \unit[520.8]{kJ}            &   \unit[1556]{kg}                &     \unit[45]{kW}          \\
4-Speed MGT                 & \unit[8.973]{MJ}      &       \unit[510.1]{kJ}            &   \unit[1561]{kg}                &     \unit[45]{kW}        \\
5-Speed MGT                 & \unit[8.988]{MJ}      &       \unit[503.6]{kJ}            &   \unit[1566]{kg}                &     \unit[45]{kW}        \\
CVT                         & \unit[9.189]{MJ}      &       \unit[499.4]{kJ}            &   \unit[1572]{kg}                &     \unit[47]{kW}        \\
2-Speed MGT with~Shift-Cost     & \unit[8.973]{MJ}    &       \unit[554.5]{kJ}             &   \unit[1551]{kg}                &     \unit[45]{kW}          \\
3-Speed MGT with~Shift-Cost     & \unit[8.976]{MJ}    &       \unit[535.6]{kJ}             &   \unit[1556]{kg}                &     \unit[45]{kW}          \\
4-Speed MGT with~Shift-Cost     &  \unit[8.990]{MJ}  &      \unit[526.8]{kJ}               &   \unit[1561]{kg}                &     \unit[45]{kW}        \\
5-Speed MGT with~Shift-Cost     &  \unit[9.009]{MJ}  &       \unit[524.0]{kJ}              &   \unit[1566]{kg}                &     \unit[45]{kW}        \\ \\

\end{tabular}
\label{tab:Results}
\end{table}

%

\section{Optimal Speed and Gear Trajectories }
\label{sec:speedandgear}
 Fig.~\ref{fig:CVT&FGT_omega&gamma} shows the optimal EM speed and gear-ratio trajectories for the EV equipped with an FGT and a CVT. The CVT is able to vary the gear ratios and therefore the range of values for the EM speed is smaller than in the case of the FGT. 
 \begin{figure}
     \centering
     \includegraphics[width = \linewidth]{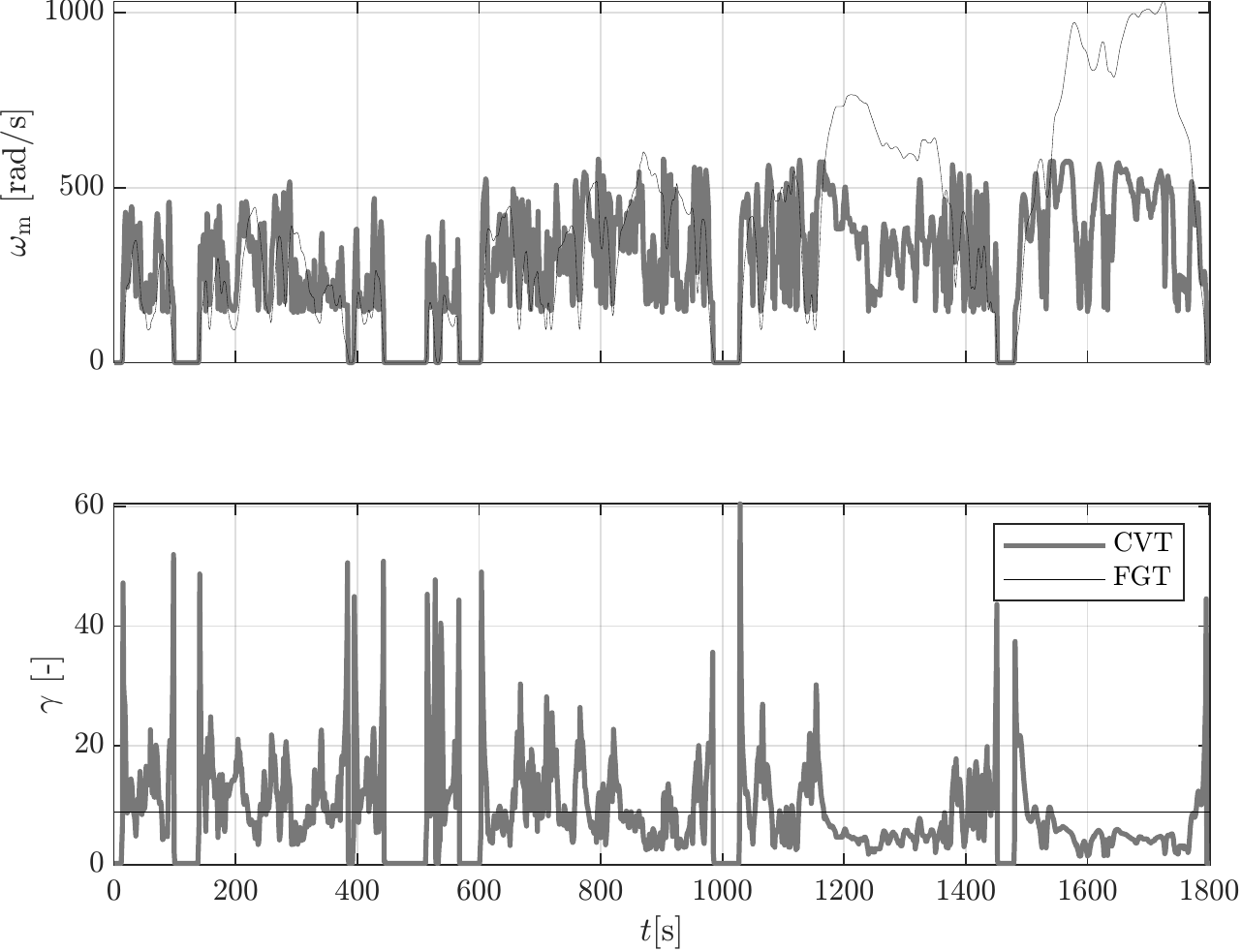}
     \caption{Optimal motor speed and gear values for an FGT and a CVT.}
     \label{fig:CVT&FGT_omega&gamma}
 \end{figure}
 Fig.~\ref{fig:23MGT_omega&gamma} and \ref{fig:45MGT_omega&gamma} show the EM speed and gear-ratio trajectories resulting from Algorithm~\ref{alg:Iterative} without gear-shifting costs.
 Going from a 2-speed to a 4-speed MGT, the increase in gears leads to a better control of the EM speed.
 In Fig.~\ref{fig:45MGT_omega&gamma}, the diminishing returns of an added gear become visible for the 4-speed and 5-speed configurations, as the difference in the EM speed becomes smaller.
 \begin{figure}
     \centering
     \includegraphics[width = \linewidth]{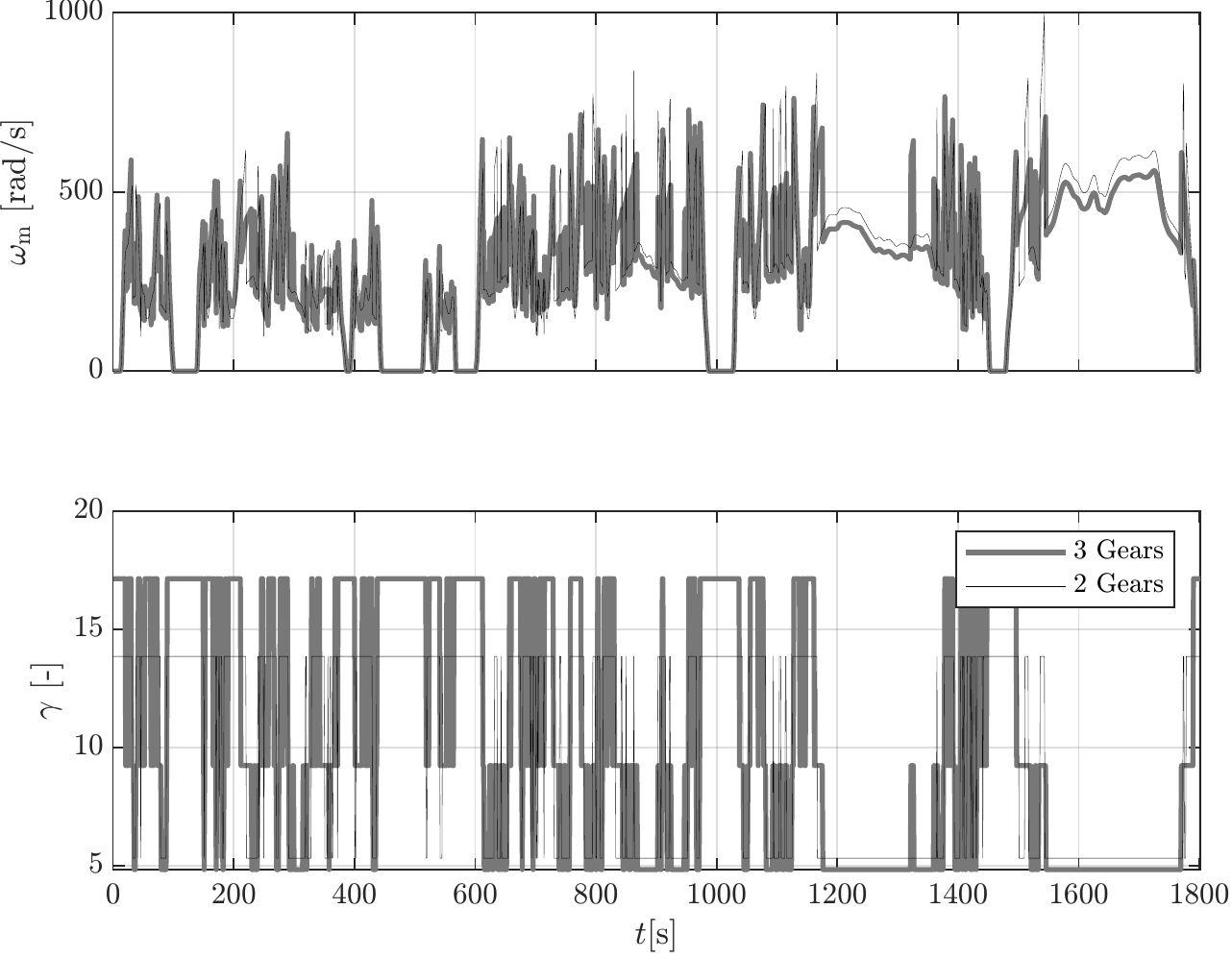}
     \caption{Optimal EM speed and gear-ratio trajectory for a 4-speed and a 5-speed MGT without gear-shifting costs.}
     \label{fig:23MGT_omega&gamma}
 \end{figure}
 \begin{figure}
     \centering
     \includegraphics[width = \linewidth]{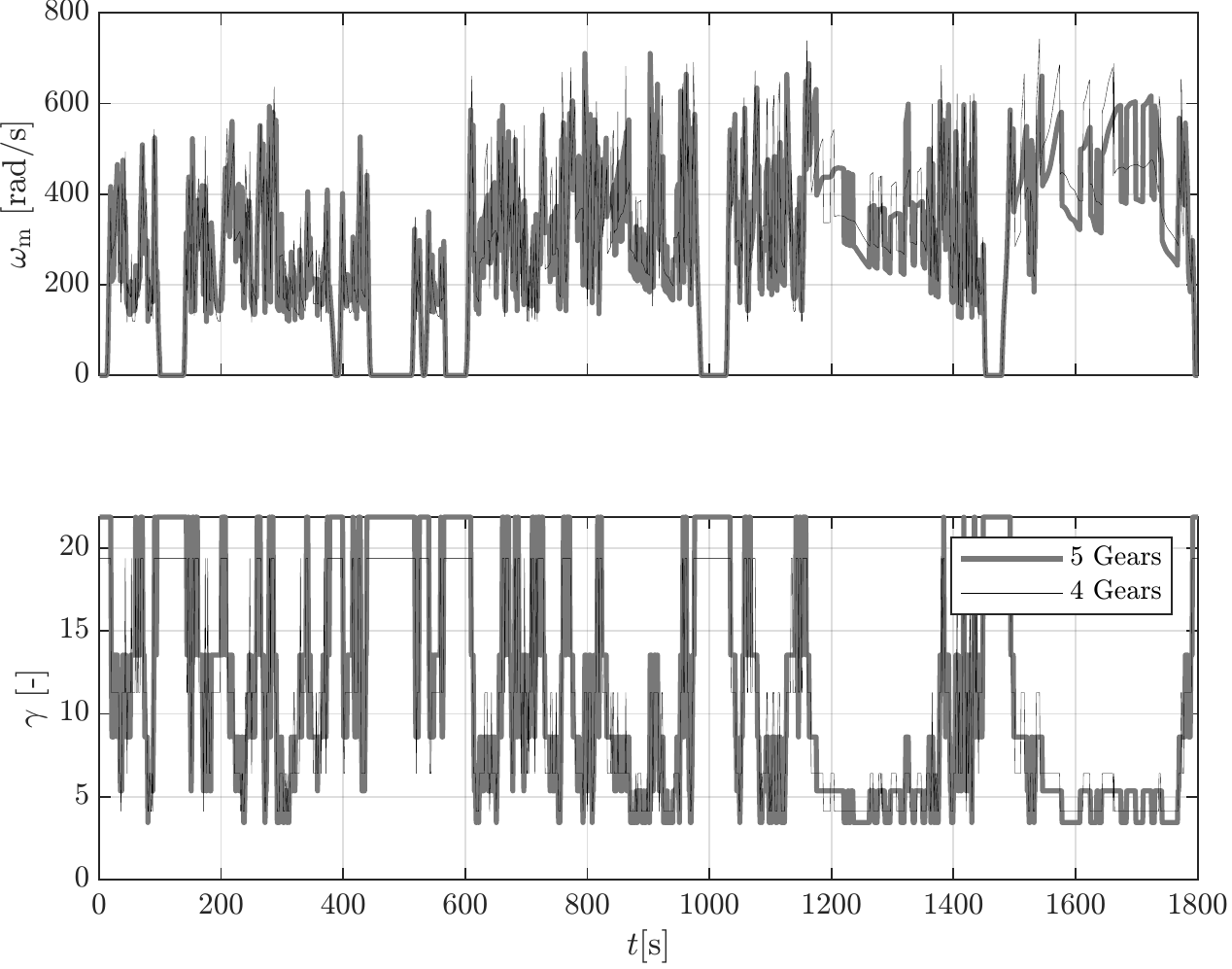}
     \caption{Optimal EM speed and gear-ratio trajectory for a 4-speed and a 5-speed MGT without gear-shifting costs.}
     \label{fig:45MGT_omega&gamma}
 \end{figure}
Fig.~\ref{fig:23MGT_omega&gamma_Gearshift_Cost} and \ref{fig:45MGT_omega&gamma_Gearshift_Cost} show the optimal EM speed and gear values for an MGT when including gear-shifting costs, whereby, compared to Fig.~\ref{fig:23MGT_omega&gamma} and \ref{fig:45MGT_omega&gamma}, we observe a clear reduction in the number of gear shifts.
 \begin{figure}
     \centering
     \includegraphics[width = \linewidth]{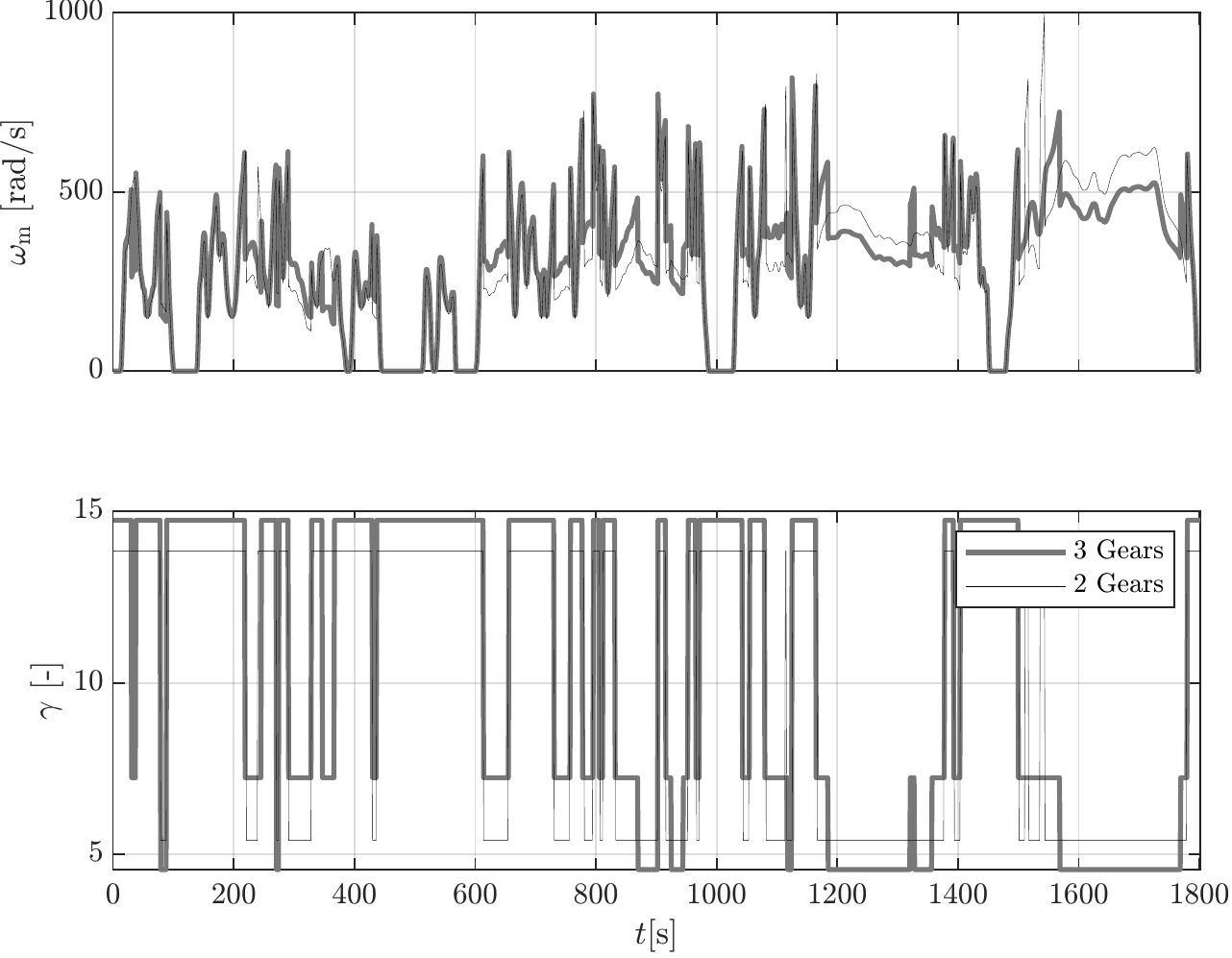}
     \caption{Optimal EM speed and gear-ratio trajectory for a 2-speed and a 3-speed MGT with gear-shifting costs.}
     \label{fig:23MGT_omega&gamma_Gearshift_Cost}
 \end{figure}
 \begin{figure}
     \centering
     \includegraphics[width = \linewidth]{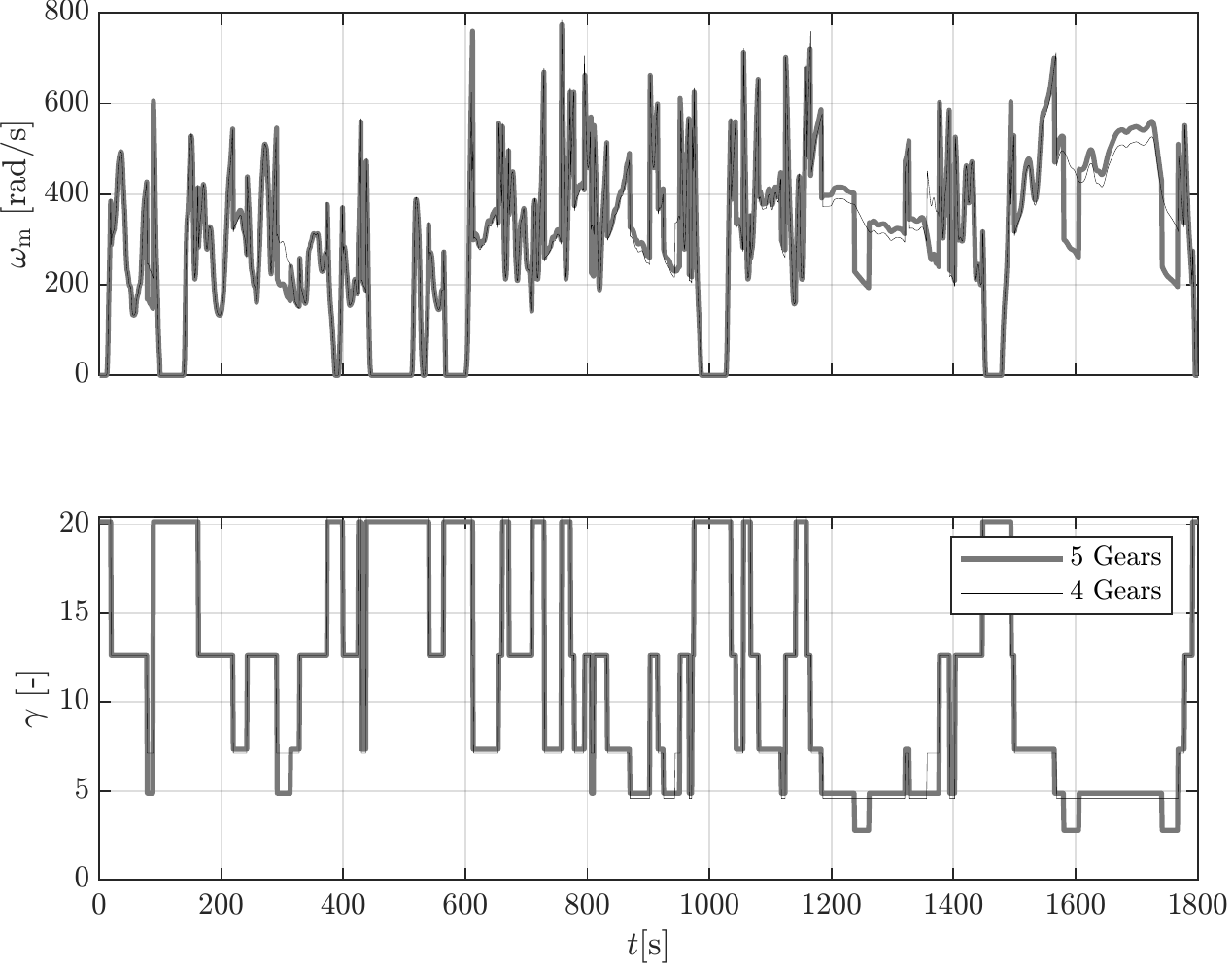}
     \caption{Optimal EM speed and gear-ratio trajectory for a 4-speed and a 5-speed MGT with gear-shifting costs.}
     \label{fig:45MGT_omega&gamma_Gearshift_Cost}
 \end{figure}
\end{appendices}
\fi
\end{document}